\documentclass[twocolumn,aps,floatfix,showpacs,bibnotes,prl,superscriptaddress]{revtex4}
\newcommand{\be}{\begin{equation}}
\newcommand{\ee}{\end{equation}}
\newcommand{\bea}{\begin{eqnarray}}
\newcommand{\eea}{\end{eqnarray}}

\usepackage{graphicx}% Include figure files
\usepackage{bm}% bold math
\begin{document}

\title{Observation of subdiffusion of a disordered interacting system}
\author{E. Lucioni}
\email{lucioni@lens.unifi.it}
\author{B. Deissler}
\author{L. Tanzi}
\author{G. Roati}
\author{M. Zaccanti}
\altaffiliation[Present address: ]{Institut f\"{u}r Quantenoptik und Quanteninformation, Innsbruck, Austria}
\affiliation{LENS and
Dipartimento di Fisica e Astronomia, Universit\`a di Firenze,
  and INO-CNR,  50019 Sesto Fiorentino, Italy }
\author{M. Modugno}
\affiliation{Department of Theoretical Physics and History of Science, UPV-EHU, 48080 Bilbao, Spain}
\affiliation{IKERBASQUE, Basque Foundation for Science, 48011 Bilbao, Spain}
\author{M. Larcher}
\author{F. Dalfovo}
\affiliation{INO-CNR BEC Center and Dipartimento di Fisica, Universit\`a di Trento, 38123 Povo, Italy}
\author{M. Inguscio}
\author{G. Modugno}
\email{modugno@lens.unifi.it}
\affiliation{LENS and
Dipartimento di Fisica e Astronomia, Universit\`a di Firenze,
  and INO-CNR,  50019 Sesto Fiorentino, Italy }

\begin{abstract}
We study the transport dynamics of matter-waves in the presence of disorder and nonlinearity. An atomic Bose-Einstein condensate that is localized in a quasiperiodic lattice in the absence of atom-atom interaction shows instead a slow expansion with a subdiffusive behavior when a controlled repulsive interaction is added. The measured features of the subdiffusion are compared to numerical simulations and a heuristic model. The observations confirm the nature of subdiffusion as interaction-assisted hopping between localized states and highlight a role of the spatial correlation of the disorder.
\end{abstract}

\date{\today}
\maketitle
The combination of disorder and nonlinearities determines the transport properties of many physical systems, including metals and superconductors \cite{Dubi07}, graphene \cite{Mucciolo10} and DNA \cite{Endres04}, or light in disordered nonlinear media \cite{Gredeskul92,Pertsch04, Schwartz07, Lahini08, Lahini09} and ultracold quantum gases \cite{Lewenstein10}. While a full understanding of the interplay of disorder and nonlinearities has long been sought, systematic experimental investigations are difficult and various aspects of this interplay have not been fully clarified. One of the open questions concerns the dynamics of a wavepacket expanding in a disordered potential in presence of a nonlinearity. There have been various predictions in theory and debate based on the results of numerical experiments over the last 20 years about this subject \cite{Shepe93, Kopidakis08, Pikovsky08, Flach09, Skokos09, Veksler09, Mulansky10, Laptyeva10, Iomin10, Larcher09}. Most authors agree that the nonlinearity should prevent localization and that the wavepacket should expand in a way which is slower than normal diffusion. However, experimental evidence of such subdiffusive expansion is to date still missing.

Here we study the dynamics of Bose-Einstein condensates with controllable nonlinearity expanding along a one-dimensional quasiperiodic lattice. Despite its large spatial correlation, this kind of potential is known \cite{Harper,Aubry} to feature exponentially localized states that are equivalent to those appearing in lattices with uncorrelated disorder described by the Anderson model \cite{Anderson58}. Such a system has been successfully exploited for the investigation of the delocalizing effect of repulsive interactions on a trapped system in equilibrium \cite{Deissler10,Deissler11}. If a non-interacting gas is let free to expand along the quasiperiodic lattice, no transport is observed \cite{Roati08} because all single-particle eigenstates are localized. By adding a controlled interatomic repulsion, we now observe a slow increase of the width $\sigma$ of the sample that asymptotically follows a subdiffusive law: $\sigma(t)\propto t^\alpha$, with $\alpha=0.2-0.4$.  We find that the exponent increases with the interaction energy, in qualitative agreement with both numerical simulations based on a 1D discrete nonlinear Schr\"{o}dinger equation (DNLSE) of the quasiperiodic lattice \cite{Larcher09} and the predictions of a heuristic model. Our observation confirms the nature of subdiffusion as interaction-assisted hopping between localized states. The observed exponents are however larger than the one calculated for uncorrelated disordered potentials \cite{Shepe93, Kopidakis08, Pikovsky08, Flach09, Skokos09, Veksler09, Mulansky10, Laptyeva10, Iomin10}, suggesting a role of the spatial correlation of the disorder.

The one-dimensional quasiperiodic potential is created by perturbing a primary optical lattice with a weaker incommensurate lattice \cite{Fallani07}:
$V(x)$=$V_1 \cos^2(k_1 x) + V_2 \cos^2(k_2 x)$. Here $k_i$=$2\pi/\lambda_i$ are the wavevectors of the lattices ($\lambda_1$=1064.4~nm and $\lambda_2$=859.6~nm). This potential is characterized by the spacing $d$=$\lambda_1/2$ and the tunneling energy $J$ of the primary lattice, and by the disorder strength $\Delta$, which scales linearly with $V_2$ \cite{Modugno09}. In the case of non-interacting particles this system constitutes an experimental realization of the Harper or Aubry-Andr\'e model \cite{Harper,Aubry} which shows a transition between extended and localized states for a finite value of the disorder $\Delta/J=2$ \cite{Modugno09,Deissler10}. Above this threshold all single-particle eigenstates of the first band of the lattice are exponentially localized, with a localization length $\xi\approx d/\ln(\Delta/2J)$.

We employ a Bose-Einstein condensate of $^{39}$K atoms in their lowest internal state, whose $s$-wave scattering length $a$ can be tuned by means of a Feshbach resonance \cite{Roati07,Derrico07}. The condensate is produced in an optical trap at $a=280 a_0$, and contains about $5\times 10^4$ atoms. We first load the interacting condensate into a quasiperiodic lattice with a constant $\Delta\approx3J$. The trap potential has a radial (axial) frequency of $ 2\pi\times50\,(70)$~Hz, while the lattice beams give an additional radial confinement with frequency $\omega_r=2\pi\times50$~Hz. At a given time $t=0$ the optical trap is suddenly switched off, allowing the sample to expand along the lattice; at the same time, $\Delta$ and $a$ are tuned to their final values within $10$~ms, and kept there for the rest of the evolution. The subsequent change of the radially-integrated spatial distribution $n(x)$ of the sample is then monitored by in-situ absorption imaging for increasing times, up to $t=10$~s. The width of the distribution is measured as the square root of its second moment: $\sigma$=$(\int x^2n(x)dx)^{1/2}$. The initial interaction energy per particle is estimated as $E_{int}$=$gN/2\int\varphi^4d^3x$, where $g$=$4\pi\hbar^2a/m$ is the coupling constant, $\varphi$ is a Gaussian approximation of the single-site wavefunction and $N$ is the mean atom number per site. We estimate that the initial distribution occupies on average 20$\pm$7 sites; this uncertainty translates into a 35\% uncertainty on $E_{int}$. Note how in this work we stay in the regime where $\xi\approx d$ is much smaller than $\sigma$.

\begin{figure}[htbp]
\includegraphics[width=0.9\columnwidth,clip]{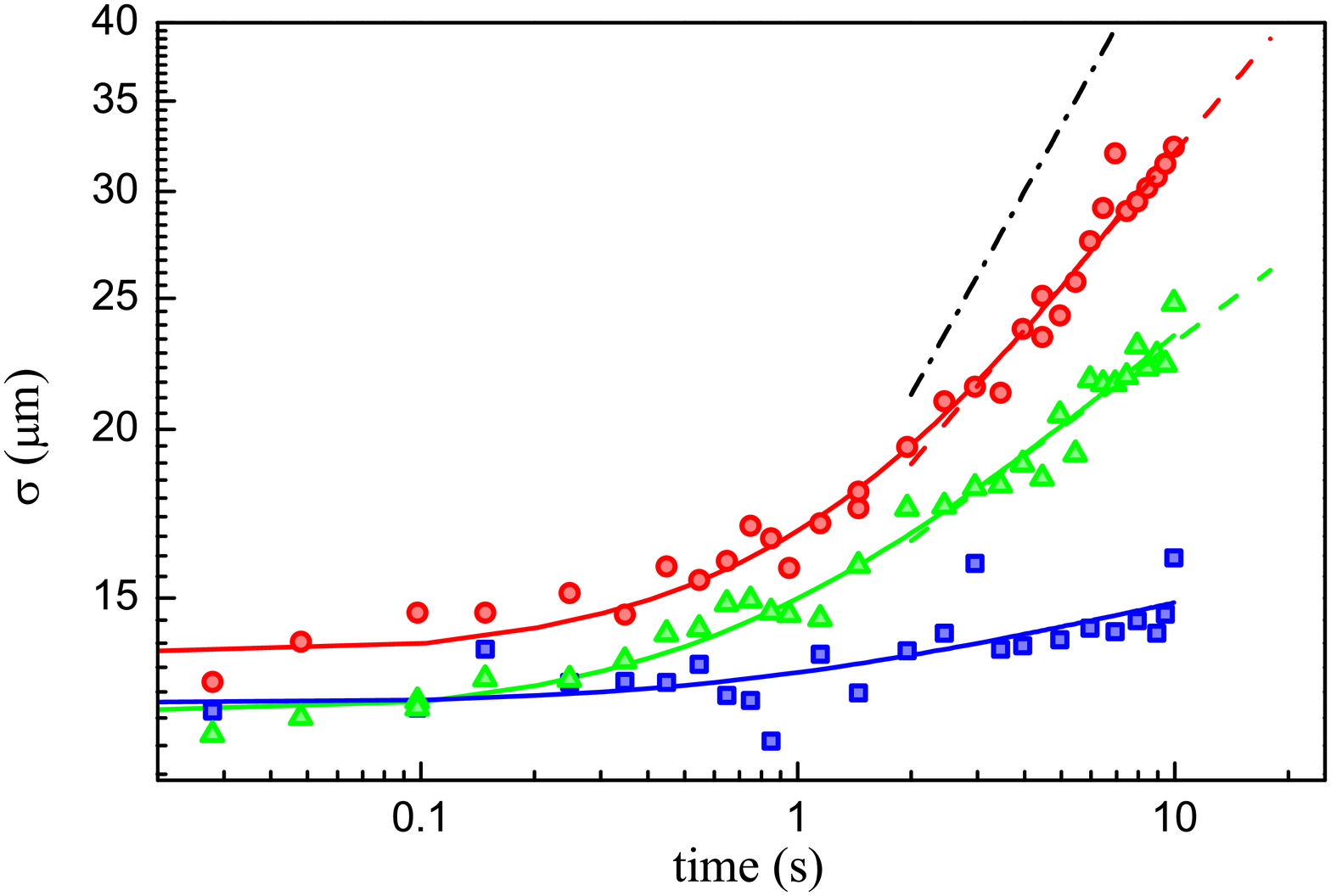}
\caption{(color online) Time evolution of the width $\sigma$ for different initial interaction energies: $E_{int}=0$ (squares), $E_{int}=1.8J$ (triangles), and $E_{int}=2.3J$ (circles). The continuous lines are the fit with Eq.(1). The dashed lines show the fitted asymptotic behavior, while the dash-dotted line shows the expected behavior for normal diffusion. The lattice parameters are $J/h=180$ Hz, $\Delta/J=4.9$.} \label{fig1}
\end{figure}

In Fig.~\ref{fig1} we present the typical time evolution of the width of the system for $\Delta/J>2$ for different interaction energies. In the absence of interaction, one observes only an extremely slow expansion, presumably due to technical noise. The introduction of a repulsive interaction allows the wave-packet to expand significantly: the expansion is however not ballistic since its velocity decreases with increasing widths. To model the expansion we fit the measured evolution with the solution of a generalized diffusion equation:
\begin{equation}
\sigma(t)=\sigma_0(1+t/t_0)^{\alpha}\,,
\label{fit}
\end{equation}
which is expected to correctly model the overall behavior from short times to the asymptotic regime \cite{EPAPS}. Here $\sigma_0$ is the initial width, $t_0$ is an "activation time" and $\alpha$ is the diffusion exponent (see also Eq.(2) later on). For normal diffusion $\alpha$=$0.5$; from the fits instead we extract $\alpha\approx0.2-0.4$, indicating subdiffusion. From the fit we typically obtain values for $t_0$ of the order of 1 s, indicating that the asymptotic behavior $t>t_0$ is achieved approximately for one decade. The expansion becomes faster when increasing the interaction energy. This is confirmed by the systematic study of the exponent reported in Fig.~\ref{fig2}, where one sees a clear increase of $\alpha$ with $E_{int}$, up to about $\alpha=0.4$.
We compare these observations with numerical simulations we have performed on a 1D DNLSE \cite{Larcher09}, for parameters close to the ones of the experiment \cite{EPAPS}. The results shown in Fig.\ref{fig2} are in qualitative agreement with the experiment: $\alpha$ increases with $E_{int}$, to approach a saturation value around $\alpha\approx0.35$. Note that noise in the experiment gives an $\alpha>0$ already for $E_{int}$=0. This bias might justify the slightly larger values of $\alpha$ in the experiment than in theory \cite{EPAPS}. There is however a disagreement on the typical energy at which the saturation regime for the exponents is reached in the experiment and in the theory, which we will discuss later.

\begin{figure}[htbp]
\includegraphics[width=0.9\columnwidth,clip]{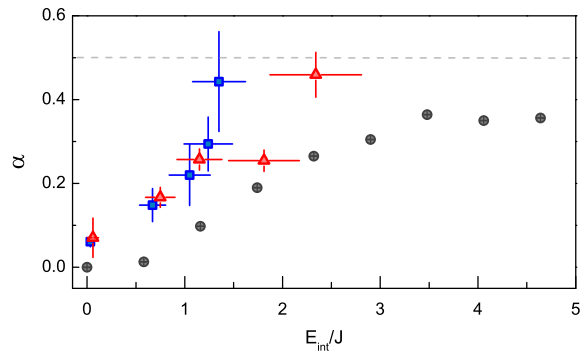}
\caption{(color online). Diffusion exponent $\alpha$ vs the initial interaction energy $E_{int}$ in the experiment (triangles and squares) and simulations (circles). The experimental data are for $\Delta/J=5.3(4)$ and two different values of the tunneling: $J/h=180$ Hz (triangles) and  $J/h=300$ Hz (squares).  The vertical bars are the fitting error of Eq.(1) to the data, while the horizontal bars indicate the statistical error. } \label{fig2}
\end{figure}

The observed subdiffusion confirms the microscopic mechanism of the expansion expected for an interacting disordered system where all the single-particle states are localized. The interaction breaks the orthogonality of the states and allows the transfer of population between neighboring states. Since the transfer rate depends on $E_{int}$, the velocity of expansion decreases as the sample expands and becomes less dense. Various authors have built heuristic models for the expansion that relate the transfer rate $\Gamma$ to the variation of the width $\sigma$ in a variety of 1D random models, such as the kicked rotor \cite{Shepe93}, the Anderson model and Klein-Gordon chains \cite{Pikovsky08, Flach09, Skokos09, Laptyeva10}. It is indeed possible to derive from basic principles, e.g. from perturbation theory, that the rate is proportional to the square of the interaction matrix elements, $\Gamma\propto |V'|^2$, with $|V'|\approx gN/2 \langle|I_{ijkl}|\rangle$ \cite{Aleiner10,EPAPS}. Here $I_{ijkl}$ is an overlap integral between four single-particle states. The average $\langle...\rangle$ is extended to all quartets of states that satisfy energy conservation, which for a disordered system amounts to have the interaction coupling term  $|V'_{ijkl}|=gN/2 |I_{ijkl}|$ larger than the energy separation between the four states \cite{Aleiner10,Altshuler97}. Note that $|V'_{ijkl}|$ is essentially $E_{int}$ times a spatial overlap coefficient that is of order unity for neighboring states.

The scaling of the atom number per site $N$ with the width, $N\propto\sigma^{-1}$, leads to an equation of motion for $\sigma$. For example, the idea followed for uncorrelated disorder \cite{Flach09,Laptyeva10} is to assume that $\Gamma$ is essentially the instantaneous diffusion constant, i.e. $d\sigma^2/dt\propto\Gamma$. The resulting equation $d\sigma/dt\propto\sigma^{-3}$ can be integrated with initial conditions $\sigma(0)=\sigma_0$ to get
\begin{equation}
\sigma(t)=\sigma_0\left(1+Ct/(\alpha\sigma_0^{1/\alpha})\right)^{\alpha}\,,\label{eq2}
\end{equation}
with $\alpha=1/4$ and $C\propto g^2$. This result holds in the regime of large interaction, where all the single-particle states within a few $\xi$ can be coupled by the interaction. Otherwise, also the number of possible couplings decreases with some power of $gN$ and the expected exponent is smaller, e.g. $\alpha=1/6$ for uncorrelated disorder\cite{Laptyeva10}.

The values of $\alpha$ we extract from the experimental data and from numerical simulations are as large as 0.35-0.4, hence definitely larger than the maximum $\alpha=0.25$ found for the uncorrelated random models above. This suggests a role of the non-decaying correlation of the quasiperiodic lattice, originating from the quasiperiodic arrangement of the single-particle energies \cite{Modugno09}. This might allow the expansion to proceed through almost coherent hopping processes, in contrast to the incoherent mechanism expected for the uncorrelated models. In this case one might relate $\Gamma$ to the instantaneous velocity, i.e. $d\sigma/dt\propto\Gamma$, which leads to Eq.(2) with an exponent $\alpha=1/3$ in case of large interaction. This regime is reached when $E_{int}$ is larger than the standard deviation of the single-particle energies $\delta E$. For $E_{int}<\delta E$, a numerical evaluation of the number of active couplings as in Ref.\cite{Flach09} shows that there is an additional scaling of $\Gamma$ approximately as $gN$ \cite{EPAPS}. Therefore in this regime one expects a smaller exponent around $\alpha=1/4$. This is consistent with what we obtain both in the experiment and in the simulations.

\begin{figure}[htbp]
\includegraphics[width=\columnwidth,clip]{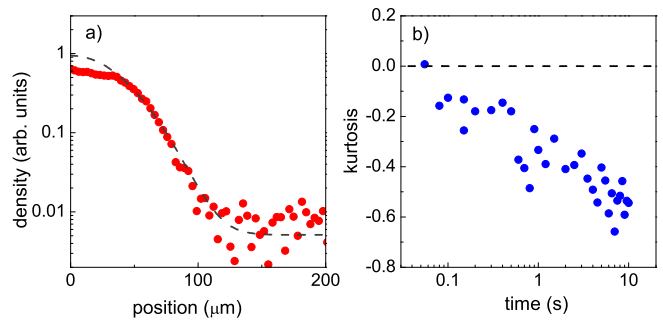}
\caption{(color online) a) Shape of an interacting sample after 10 s of evolution; the dashed curve is a gaussian fit of the tails. b) Time evolution of the kurtosis $\gamma$; the dashed line marks the kurtosis of a gaussian distribution. The interaction energy is $E_{int}\approx1.8J$, all other parameters as in Fig.\ref{fig2}.} \label{fig3}
\end{figure}

Since the coupling between localized states is largest at the center of the sample, where the atom number per site is largest, one expects a faster expansion at the center than in the tails \cite{Shepe93}. This behaviour is actually clearly visible in the typical shape of the cloud at a long $t$ shown in Fig.~\ref{fig3}a, which features an extended flat top and rapidly decaying tails. The flat-top shape gradually emerges from initially gaussian profiles during the expansion, as can be seen in the typical evolution of the kurtosis, $\gamma=(\int x^4 n(x)dx)/\sigma^4 - 3$, in Fig.~\ref{fig3}b. A comparison with general models of nonlinear diffusion \cite{Mulansky10,Laptyeva10} is due in future studies.

An additional feature of the experimental system is the presence of the radial degrees of freedom, something that is absent in all numerical simulations and models developed so far. In our case the radial quantum $\hbar \omega_r\approx J/5$ is about one order of magnitude smaller than the typical $\delta E\approx3J$. Therefore several radial states might be populated in presence of the interaction, which couples the various degrees of freedom. The radial excitation can be taken into account by including the radial wavefunctions in the overlap integrals $I$ \cite{EPAPS}. An increasing radial excitation can change the microscopic hopping dynamics in two ways: i) It increases the radial width of the sample, and therefore decreases $I$, with a consequent slowing down of the expansion and a possible reduction of $\alpha$. ii) It provides a smaller energy scale for the coupling of the localized states, since the axial energy mismatch can now be reduced in presence of a radial excitation. Indeed, while the minimum energy mismatch in 1D is of the order of $\delta E$, it becomes of the order of $\hbar\omega_r$ for large temperature samples in 3D. This effect tends to increase the overlap $I$, especially in the case of a small $E_{int}$ that would not be able to couple efficiently states in 1D.

To check the relative magnitude of these two competing effects we have compared the evolution of interacting samples at low radial temperature $T_r$ with those of samples with the same $g$, but with a larger $T_r$ prepared via controlled parametric heating. The typical observation is reported in Fig.~\ref{fig4}: the expansion of the high-$T_r$ sample is faster, despite a reduction of $E_{int}$ by approximately a factor 2, indicating that the second effect dominates. Raising the temperature therefore helps the coupling of a larger number of states by the interaction. In order to separate out the effect of the reduced $E_{int}$ at the larger $T_r$, we have also measured the time evolution of the sample at low $T_r$ in which a similar $E_{int}$ was achieved by reducing $g$. This measurement shows a slower expansion, as expected since now the reduced $E_{int}$ is not compensated by an increased coupling efficiency. The effect of the radial degrees of freedom can also explain the disagreement in the energy scales between theory and experiment in Fig.\ref{fig2}: in the 1D simulation the relevant energy for being in the large interaction regime is $\delta E\approx 3J$, while in the experiment in 3D this is shifted towards $\hbar\omega_r\approx J/5$.

From a study of the radial momentum distribution of the interacting samples we also detect a radial heating during the 10~s evolution time, with a typical rate of the order of 3-5~nK/s that is at least twice as large as the heating of non-interacting samples \cite{EPAPS}. This heating leads to final $T_r$ that can be larger than $\delta E/k_B\approx$~50~nK, so that it cannot be entirely justified by thermalization of the axial and radial degrees of freedom. We attribute such excess heating to energy transfer from the axial noise mentioned above to the radial degrees of freedom, mediated by the interaction. Finally, we note that for non-interacting samples we do not observe any dependence of the axial dynamics on the radial temperature, confirming the secondary role of the radial modes in the subdiffusion.

\begin{figure}[htbp]
\includegraphics[width=0.9\columnwidth,clip]{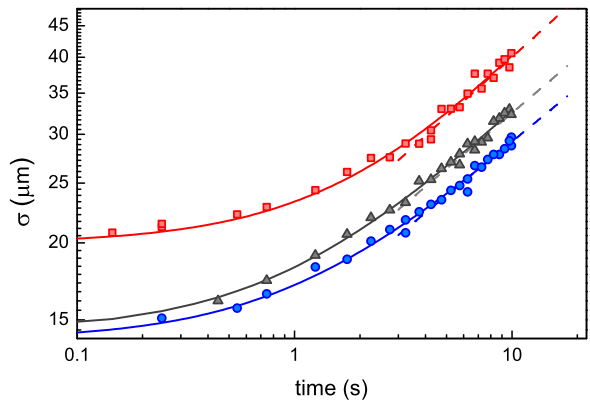}
\caption{(color online) Effect of a finite radial temperature. The scattering length and mean temperature of the three data-sets are: $a=100a_0$, $T_r=60$nK (circles); $a=400a_0$, $T_r=60$nK (triangles); $a=400a_0$, $T_r=200$nK (squares). The continuous lines are the fit with Eq.(1), while the dashed lines are the fit to the asymptotic behavior. The fitted exponents are: $\alpha$=0.32(2) (circles); $\alpha$=0.34(2) (triangles); $\alpha$=0.40(5) (squares). The lattice parameters are $J/h$=290 Hz, $\Delta/J$=3.9.} \label{fig4}
\end{figure}

In conclusion, our study gives evidence of the subdiffusive character of the expansion of an interacting system in a disordered potential. On the timescales considered here, it qualitatively validates the existing models of the subdiffusion as interaction-assisted hopping between localized states, although it leaves open the question about the fate of the expansion at very long times \cite{Laptyeva10,Zhang}. Our study raises new questions about the role of the correlations of the potential, which would be interesting to explore also in different kinds of disorder with intermediate correlations. Also, we could drive our system into the quasi-1D regime by increasing the radial confinement to freeze out the radial degrees of freedom. There one could study the predicted temperature-induced metal-insulator transition \cite{Aleiner10}, or strongly correlated regimes \cite{Roux08,Radic10}.

We acknowledge discussions with F. Caruso, C. D'Errico, A. M. Garcia-Garcia, M. Moratti, S. Flach, F. M. Izrailev, S. Lepri, A. Pikovsky, D. Shepeliansky. This work was supported by EU (IP AQUTE), by ERC through the QUPOL grant and by the ESF and CNR through the EuroQUASAR program.

\section{Supplementary material}

\section{A perturbative model of subdiffusion}

In this section we discuss with some more detail the process of hopping between localized states induced by the interaction that gives rise to subdiffusion. A modeling of the subdiffusion of disordered, interacting systems has already been given by several authors \cite{Shepe93,Flach09,Laptyeva10} by analyzing the properties of the Discrete Nonlinear Schr\"{o}dinger Equation (DNLSE). Here we present an analogous heuristic model based on perturbation theory concepts. Although this model cannot be used to describe quantitatively the full dynamics, it provides correct estimates for the subdiffusion exponents. Let us start by considering the localized eigenstates of the 1D single-particle problem, $\varphi_i$, labeled by the site index $i$. These states have individual energies $\epsilon_i$ with typical separation $\delta E\approx\Delta$ and a characteristic localization length $\xi\approx d/\ln(\Delta/2J)$.
\begin{figure}[htbp]
\includegraphics[width=\columnwidth,clip]{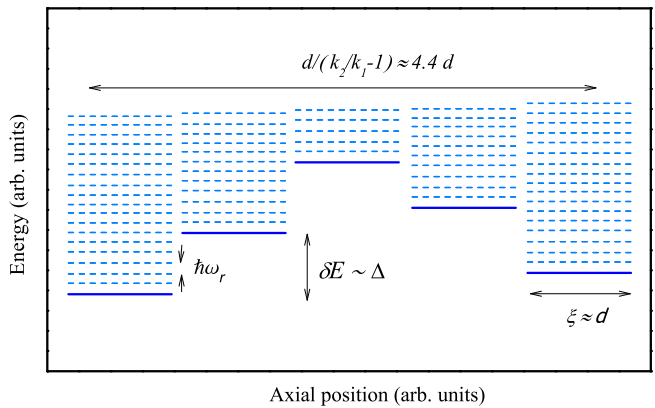}
\caption{Cartoon of the energy spectrum of the axial states along one quasiperiod of the quasiperiodic lattice (continuous lines) together with their associated radial states (dashed lines). The localization length $\xi$ is of the order of the lattice spacing $d$.} \label{fig1}
\end{figure}
The interaction Hamiltonian operator is
\be
\hat{H'}=\frac{g}{2} \int\hat{\psi}^{\dag}(x)\hat{\psi}^{\dag}(x)\hat{\psi}(x)\hat{\psi}(x) dx\,,
\ee
where
\be
g=\frac{4\pi\hbar^2}{m}a\,,\,\,\,\,\,\hat{\psi}(x)=\sum_i  \hat{b}_i\varphi_i(x)\,,
\ee
with $\hat{b}_i$ the bosonic destruction operator at site $i$. In general, up to four states can be coupled by the interaction. The dominant process is the one in which two particles are moved from the states $i$ and $j$ with occupation $N$ to two initially empty states $k$ and $l$. Let us indicate by $N$ the mean number of atoms per site in the initial distribution, which is related to the width in units of lattice sites $\sigma/d$ by the relation $N=N_{tot} d/\sigma$. The off-diagonal term of the interaction Hamiltonian for this process has the form
\be
V'_{ijkl}=\frac{g}{2}N\int\varphi_l^{\ast}(x)\varphi_k^{\ast}(x)\varphi_j(x)\varphi_i(x) dx\,.
\label{off}
\ee
From perturbation theory we know that this process can happen at asymptotic times only if the energy is conserved, i.e. $\Delta E=|\epsilon_i+\epsilon_j-\epsilon_k-\epsilon_l|=0$. In a disordered system it is impossible to have perfect energy matching of the states, but on the other hand if the transition rate for the process is large enough, even a finite energy difference cannot be resolved and the transition can take place. As discussed for example in Refs.~\cite{Aleiner10,Altshuler97} the appropriate energy conservation requirement is actually
\be
|V'_{ijkl}|>\Delta E\,.
\ee
The transfer rate associated to this microscopic process is
\be
\Gamma_{ijkl}=\frac{2\pi}{\hbar}\frac{|V'_{ijkl}|^2}{\Delta E}\,.
\ee
From Eq.~3 one sees that the coupling term $|V'_{ijkl}|$ is essentially the interaction energy $E_{int}$ times an overlap integral $I\approx\exp(-L/\xi)$, where $L$ is the mean separation between the four states. We can therefore conclude that the macroscopic expansion of the system will be determined by all the microscopic processes between states that are laying within a few localization lengths from each other.
It is instructive to numerically calculate the typical coupling strength in the quasiperiodic lattice for transitions between doublets of states, $|V'_{iikk}|/\Delta E_{ik}$, for various separations between the two states. As shown in Fig.~\ref{fig2}, the largest coupling is achieved on average for states that are separated by one quasiperiod of the lattice, $d/(k_2/k_1-1)$. This result is a consequence of the peculiar spatial correlation of the eigenstates in the quasiperiodic lattice.

\begin{figure}[htbp]
\includegraphics[width=0.9\columnwidth,clip]{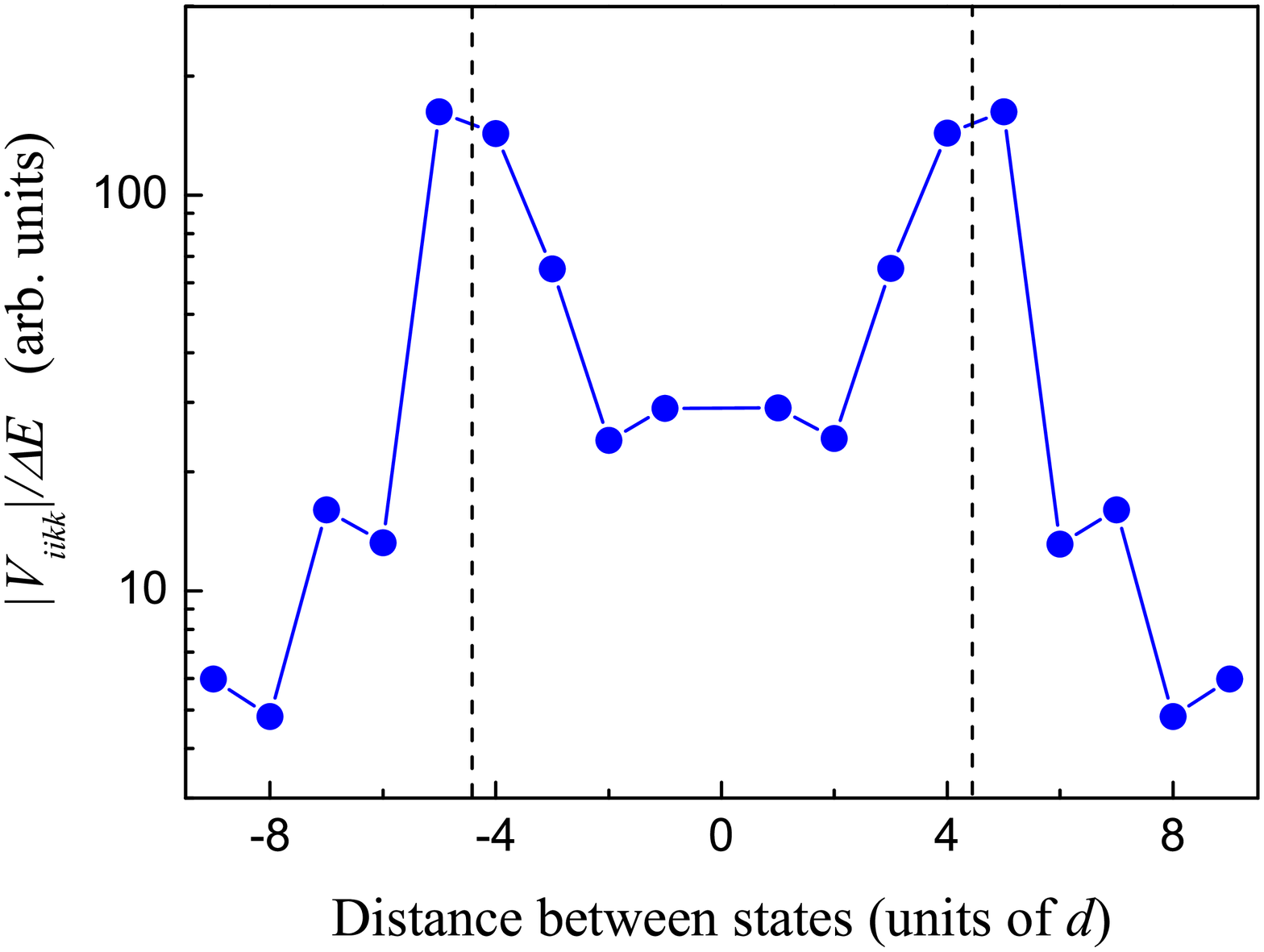}
\caption{Calculated mean coupling for doublets of states versus the distances between the states for a quasiperiodic lattice with $\Delta=4J$. The dominant couplings are those to states about one quasiperiod, $d/(k_2/k_1-1)\approx 4.4d$, apart.} \label{fig2}
\end{figure}

One can distinguish two different regimes of expansion, along the lines of the discussion in Ref.~\cite{Laptyeva10} and references therein.

1) If the initial interaction energy is sufficiently large, then at least the strongest couplings of each localized state to the other states within one localization length are active. This regime is reached when $E_{int}>\delta E$. Here one can define a macroscopic rate $\Gamma=\langle\Gamma_{ijkl}\rangle$, where the symbol $\langle...\rangle$ denotes an average over the system extension. From Eqs.~3 and 5 one obtains $\Gamma\propto(gN)^2\propto\sigma^{-2}$. We can now relate $\Gamma$ to a rate of change of the width of the distribution $\sigma$. In the case in which the expansion is assumed to be instantaneously diffusive, as done for example in \cite{Flach09} to describe a lattice with random disorder, the diffusion rate is essentially $\Gamma$, and one obtains
\be
d\sigma^2/dt\propto\sigma^{-2}\,.
\ee
The large coupling of states separated by one quasiperiod shown in Fig.~\ref{fig2} suggests however that in our quasiperiodic lattice the expansion might have a more coherent nature, i.e. the expansion is dominated by sequential hopping between states close in energy and separated by approximately one quasiperiod. In the case of such an expansion mechanism, the rate $\Gamma$ represents instead the instantaneous velocity, and one obtains
\be
d\sigma/dt\propto\sigma^{-2}\,.
\ee

2) If instead $E_{int}<\delta E$, only a reduced number of microscopic transitions are activated by the interaction, and the rate $\Gamma$ has an additional dependence on $E_{int}$. For example, we numerically evaluated the behavior of the couplings between doublets of states for our quasiperiodic lattice with $\Delta=3.5J$ (which corresponds to $\delta E\approx 3J$), following the analysis presented in Refs.~\cite{Flach09,Flach10}. To do this, we selected the strongest coupling $|V'_{iikk}|/\Delta E_{ik}$ for each initial state $i$ and we calculated the probability to have such coupling active for a varying interaction energy. We have then averaged this result over many initial states along the lattice. As shown in Fig.~\ref{fig3}, we find that the probability is about constant for $E_{int}\gtrsim\delta E$, while it decreases at least as $\sqrt{E_{int}}$ for decreasing interaction energy. Since $E_{int}\propto\sigma^{-1}$, this implies that the related rate $\Gamma\propto\langle |V'_{iikk}|^2\rangle$ can scale at least as $\sigma^{-3}$. This changes the overall exponent in the equation of motion for $\sigma$. For example, in the case of a coherent expansion one obtains
\be
d\sigma/dt\propto\sigma^{-3}\,.
\ee
This regime of slower subdiffusion is analogous to the one reported in Ref.~\cite{Flach09} for uncorrelated disorder and dubbed "weak chaos" regime, in contrast to the opposite regime of "strong chaos".

\begin{figure}[htbp]
\includegraphics[width=0.9\columnwidth,clip]{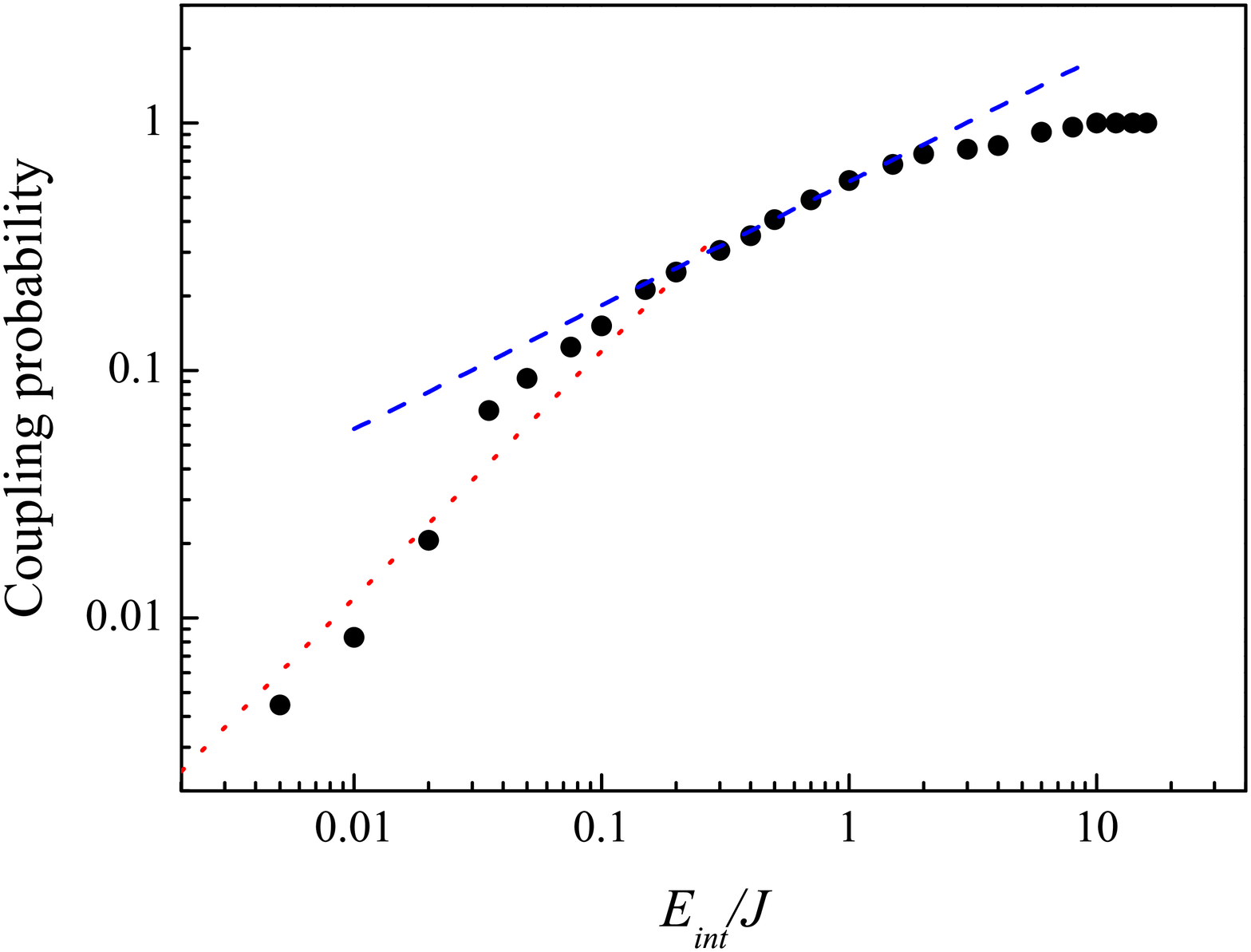}
\caption{Calculated probability of coupling for doublets of states in a quasiperiodic lattice with $\Delta=3.5J$ (see text). The dashed and dotted lines are a guide to the eye for decays proportional to $\sqrt{E_{int}}$ and $E_{int}$, respectively.} \label{fig3}
\end{figure}

\section{Fitting the axial dynamics in the experiment and in the numerical simulations}

Let us now discuss in more detail the link between the microscopic model and the macroscopic expansion observed in the experiment and in the numerical simulations. We have derived above an equation of motion for the width of the form: $d\sigma/dt=C\sigma^{-p}$, where $p$ is a positive rational number and $C$ is a constant which takes in account all the terms that do not depend directly on $\sigma$. The integration of this equation with the initial condition $\sigma(0)=\sigma_0$ leads to
\begin{equation}
\sigma(t)=\sigma_0\left(1+\frac{C}{\alpha\sigma_0^{1/\alpha}} t\right)^{\alpha}\,,
\label{fit}
\end{equation}
with $\alpha=1/(p+1)$. This can also be expressed as
\be
\sigma(t)=\sigma_0\left(1+\frac{t}{t_0}\right)^{\alpha}\,,
\ee
where the parameter $t_0$ represents an "activation time". In general, three fitting parameters are therefore required to fit the data. For $t\gg t_0$ the initial size does not play a role in the expansion, two parameters in the fitting procedure are enough and the expansions shows a linear behavior in the log-log scale. However, in order to decrease the uncertainty on the parameters and to extract information also from the measurements at shorter times, we chose to fit also the data taken for $t<t_0$ using the three parameters fitting function. This gives also the possibility to confirm that the asymptotic regime has been reached by comparing $t_0$ with the maximum observation time. The characteristic value for the activation time $t_0$ we fitted for the sets of data reported in the paper is indeed around one second (except for the non-interacting case, where it is larger), i.e. one order of magnitude smaller than the maximum observation time. Note also that by fitting only the data points for $t>t_0$ with the asymptotic form of Eq.~\ref{fit} we obtain exponents $\alpha_{lin}$ which are in good agreement with the previous ones. One example of this comparison can be seen in Fig.~\ref{fig4}a.

\begin{figure}[htbp]
\includegraphics[width=0.9\columnwidth,clip]{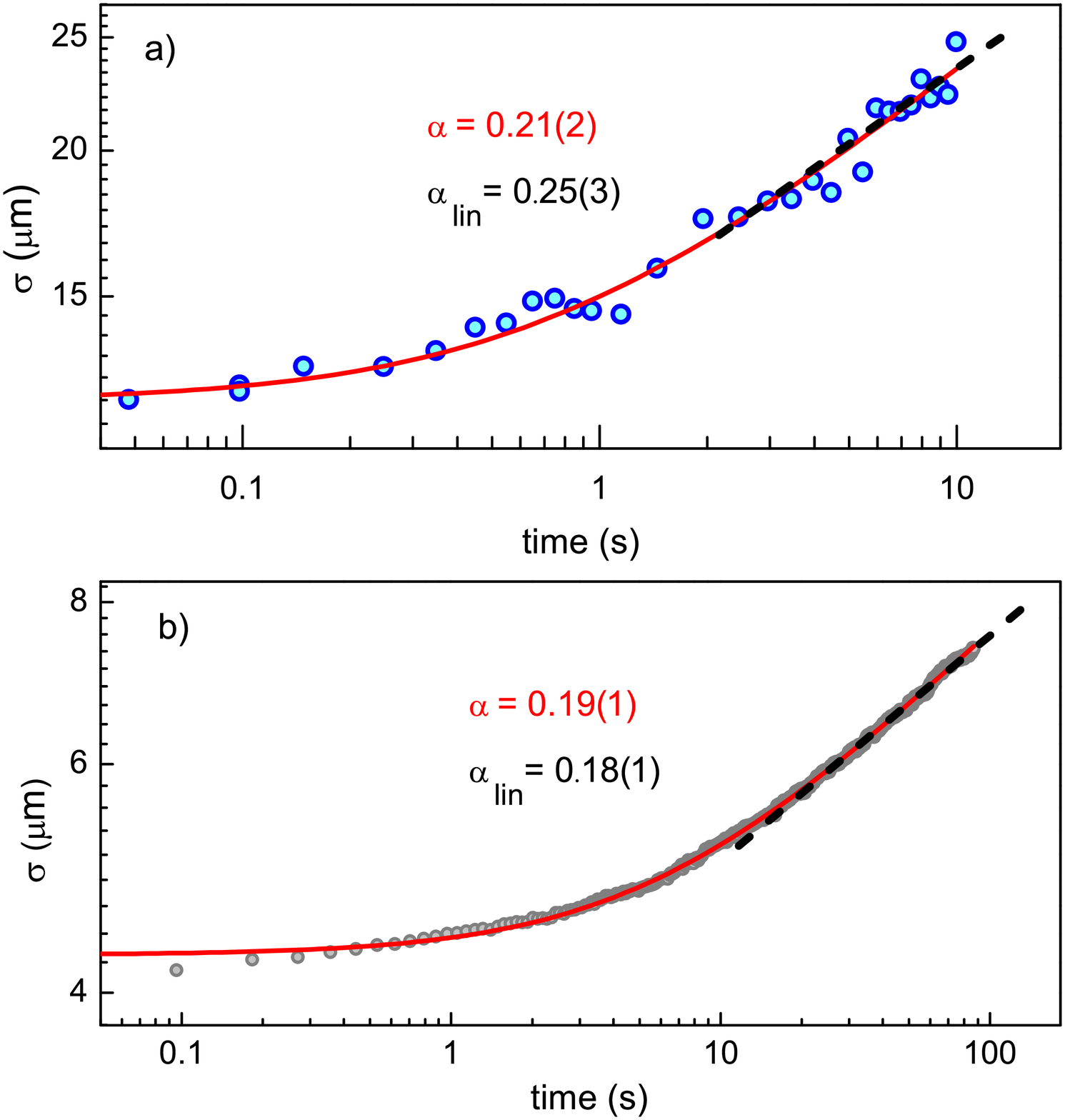}
\caption{Typical time evolution of the axial width $\sigma$ for an interacting sample. The value for the diffusion exponent from a linear fit in log-log scale for asymptotic times $\alpha_{lin}$ (black dashed line) is in good agreement with the one from the three parameters fitting function $\alpha$ (red line). Panel a) showns the experimental data for $E_{int}\approx 1.8$; panel b) the numerical simulation for an interaction parameter $\beta=60$ which corresponds to $E_{int}\approx 1.8$} \label{fig4}
\end{figure}

We also numerically solved the discrete 1D DNLSE, Eq.~2 of Ref.~\cite{Larcher09} using the same parameters for the lattice as in the experiment. The interaction parameter $\beta$ of Ref.~\cite{Larcher09} is related to the interaction energy $E_{int}$ of this work by $E_{int}=(\beta/2) \sum_i |\psi_ i|^2$, where $|\psi_i|^2$ is the probability of finding an atom at the lattice site $i$. We report a typical curve obtained from a simulation up to 100 seconds in Fig. 4b. As already done with the experimental data, we extracted the exponent $\alpha$ by fitting the numerical data with the expression in Eq.~\ref{fit}. We also checked that the exponent $\alpha_{lin}$ extracted by a fit of the data in the asymptotic regime alone is very close to $\alpha$.

Note that in the experiment we measure $\alpha\approx0.06$ even for $E_{int}\approx0$. This is presumably due to a weak technical noise on the quasiperiodic potential (e.g. vibrations in the retroreflecting mirror that creates the standing waves, or fluctuations of the laser frequencies). If we would assume that this value represents a constant bias for all measurements with interaction, then we would find experimental values for $\alpha$ that are closer to the theory. However, in literature there are no studies on the interplay of noise and interaction in a disordered system that can fully justify this assumption. A study that we are currently performing on our system indicates that the diffusion exponents due noise and interactions roughly add up in the regimes explored in this work, but a conclusive analysis is still missing. We therefore leave a more detailed discussion of this point to a future study.

\section{Modeling the radial degrees of freedom and temperature evolution in the experiment}

The presence of the radial degrees of freedom in the experiment can now be taken into account. As shown in Fig.~1, the radial energy spacing $\hbar \omega_r$ is about one order of magnitude smaller than the characteristic disorder energy $\delta E$. There are two main effects of the radial excitations to be considered:

1) The 4-states overlap integral must be modified to include also the radial plane
\begin{equation}
I_{ijkl}=\int \varphi_i^{\ast}\varphi_j^{\ast}\varphi_k\varphi_l dx \int \varphi_{ri}^{\ast}\varphi_{rj}^{\ast}\varphi_{rk}\varphi_{rl} d^2r\,.
\end{equation}
An increasing radial excitation of the system therefore results in a broadening of the radial wavefunctions, with a consequent decrease of the overlap integral and the associated coupling term $|V'_{ijkl}|$.

2) The possibility of populating excited radial modes softens the energy conservation requirement, since the radial level separation $\hbar \omega_r$ is smaller than the characteristic energy spacing of the quasiperiodic lattice $\delta E$ (by approximately one order of magnitude in the present experiment). This implies that axial states that could not be coupled because of a too small $E_{int}$ in a 1D system, might be at least partially coupled in presence of a radial excitation.

We have not attempted so far a quantitative numerical study of this model including the radial modes. However, the considerations above can be used to derive qualitative expectations that can explain the experimental observations. We can indeed conceive two possible limits for the radial temperature. The first one is the limit $k_B T_r<\hbar\omega_r$, which in our case ($\hbar\omega_r / k_B\approx$ 5~nK) would correspond to the minimum temperature achievable in the Bose-Einstein condensate. In this limit the excited radial population is negligible, and one has effectively a 1D problem. However, this limit cannot be achieved as a stationary state, since the thermalization of the axial and radial degrees of freedom will rapidly bring the radial temperature to a minimum value of the order of $T_r\approx \delta E/k_b\approx$ 50~nK.
The other limit is $k_B T_r\gg\delta E$, in which many radial states are excited. In this limit the energy mismatch that must be compensated by the coupling terms to provide the axial hopping is no longer of order $\delta E$, but it becomes of order $\hbar \omega_r$. The critical $E_{int}$ to reach the regime of strong chaos will therefore be $\hbar\omega_r$.
The experiment is performed in an intermediate regime of temperature, $k_B T_r\approx\delta E$, where we can expect that the critical energy for the strongly interacting regime lays somewhere in between $\delta E$ and $\hbar\omega_r$.
\begin{figure}[htbp]
\includegraphics[width=0.9\columnwidth]{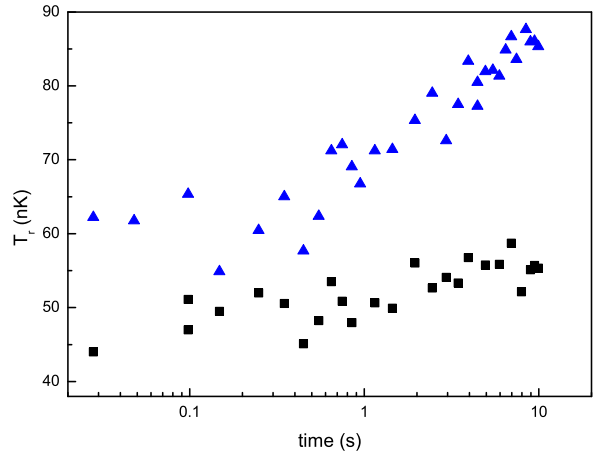}
\caption{Typical time evolution of the radial temperature for a non-interacting sample (black squares) and an interacting one (blue triangles).} \label{fig5}
\end{figure}

This expectation is confirmed by the comparison between the experimental and numerical energy dependence of the exponent $\alpha$ discussed in the paper. Note that the maximum temperature that can be reached in the experiment while preserving a single-band dynamics is of the order of 200nK. For higher temperatures we see a clear excitation of the second band of the quasiperiodic lattice, whose single-particle eigenstates are not localized for the range of parameters we explored. This excitation is detected as the appearance of rapidly moving tails in the density distribution.

Finally, we discuss the evolution of the radial temperature $T_r$ in the experiment for samples that are initially prepared as Bose-Einstein condensates, i.e. at the lowest measurable temperatures, $T_r\lesssim$ 50nK. The typical evolution is reported in Fig.~\ref{fig5}. In absence of interaction we see a slow heating with a rate of the order of 1-3 nK/s, presumably due to pointing or amplitude noise on the lattice beams. The interacting samples show instead a larger initial $T_r\approx\delta E/k_B$, which presumably arises from a thermalization of the axial and radial degrees of freedom. The following heating during the full evolution time is typically 3-5 nK/s, hence larger than the one of non-interacting samples. This excess heating presumably arises from the same axial noise on the quasiperiodic lattice that causes the slow expansion of non-interacting samples, see Fig.~1 in the paper. This noise might be due to variations in the laser wavelengths and to vibrations of the retro-reflecting mirror that creates the standing waves, which result in a spatial movement of the two lattices and a consequent excitation of the axial dynamics even in the case of orthogonal localized states. In presence of interaction the axial excitation is transferred to the radial degrees of freedom, resulting in an increased heating.

\end{document}